\documentstyle[times,pramana,epsf,floats]{ias}
\begin{document}
\mark{{Studies of Radiative Penguin B Decays}{John M. LoSecco}}
\title{Studies of Radiative Penguin B Decays at BaBar}
\author{John M. LoSecco}
\address{University of Notre Dame, Notre Dame, Indiana 46556, USA,
For the BaBar Collaboration}
\abstract{
We summarize results on a number of observations of penguin dominated radiative
decays of the B meson.  Such decays are forbidden at tree level and proceed
via electroweak loops.  As such they may be sensitive to physics beyond
the standard model.  The observations have been made at the BaBar experiment
at PEP-II, the asymmetric B factory at SLAC.}
\maketitle
\section{Introduction}
\begin{figure}[htbp]
\centerline{\epsfysize=2cm\epsfxsize=7cm\epsfbox{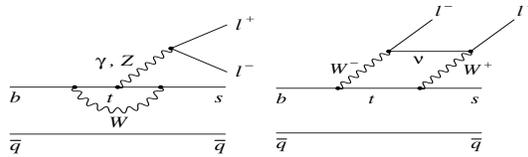}}
%\centerline{\epsfysize=2cm\epsfbox{Kll_SUSY_Diagrams.eps}}
\caption{Standard model contributions to penguin dominated decays.}
% (top) and
%potential new physics contributions to radiative decays (bottom).  The lower
%diagrams include contributions from squarks, charginos, neutralinos and
%gluinos.}
\label{fig:penguin}
\end{figure}
Penguin dominated B decays are decays which proceed via higher order
electroweak loops as illustrated in figure \ref{fig:penguin}.  Such decays
are sensitive to new
physics beyond the standard model since the virtual new particles would
contribute via loops to graphs that would
produce amplitudes that would interfere with the standard
mechanism.
\section{$B \rightarrow K^{(*)} l^{+} l^{-}$}
The decay mode $B \rightarrow K^{(*)} l^{+} l^{-}$ is a flavor changing
neutral current which, in the standard model, is mediated by a class of graphs
illustrated in figure \ref{fig:penguin}.  The search included both charged and
neutral B mesons, $K$ and $K^{*}$ final states and both electron and muon
pairs.

Predictions for this decay rate\cite{kll} are
about $5 \times 10^{-7}$ for decays to $K$ and
$2 \times 10^{-6}$ for the decays to $K^{*}$.  The analysis removes standard
model contributions coming from $B \rightarrow K^{(*)}$ $J/\psi$, $\psi'$,
%\jpsi, \psiprpr, \Psi$, $\Psi'$,
or $\gamma$.

A standard analysis method for B decays at BaBar utilizes the two kinematic
variables the beam constrained mass, $M_{ES}$ and $\Delta E$:
$M_{ES}= \sqrt{(E^{*}_{beam})^{2}-(p^{*}_{B})^{2}}$,
$\Delta E= (E^{*}_{B} - E^{*}_{beam})$.
The ${*}$ indicates that the quantiity is calculated in the CM frame (or
$\Upsilon$(4S) rest frame).
The rates measured were \cite{kll}:
$\cal B$( $B \rightarrow K l^{+} l^{-}$ ) = $(0.78^{+0.24 +0.11}_{-0.20 -0.18})
\times 10^{-6}$
(4.4 $\sigma$ including systematic),
$\cal B$( $B \rightarrow K^{*} l^{+} l^{-}$ ) = $(1.68^{+0.68}_{-0.58}\pm0.28)
\times 10^{-6}$
(2.8 $\sigma$ including systematic)
or $\cal B$( $B \rightarrow K^{*} l^{+} l^{-}$ ) $<$ $3.0 \times 10^{-6}$
%\vspace{-1cm}
\begin{figure}[htbp]
\centerline{\epsfysize=3cm\epsfxsize=5cm\epsfbox{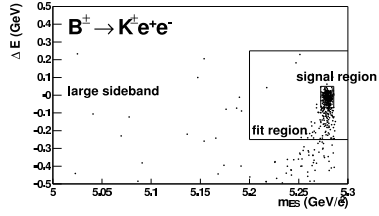}
}
\caption{Expected $M_{ES}$ and $\Delta E$ distribution from simulated
$B^{\pm} \rightarrow K^{\pm} e^{+} e^{-}$}
% (left) and
%a diagram contributing to the decay $B \rightarrow K^{*} \gamma$ (right)}
\label{fig:kllmc}
\end{figure}
\section{$B \rightarrow K^{*} \gamma$}
%\vspace{-1cm}
\begin{figure}[htbp]
\centerline{\epsfysize=5cm\epsfbox{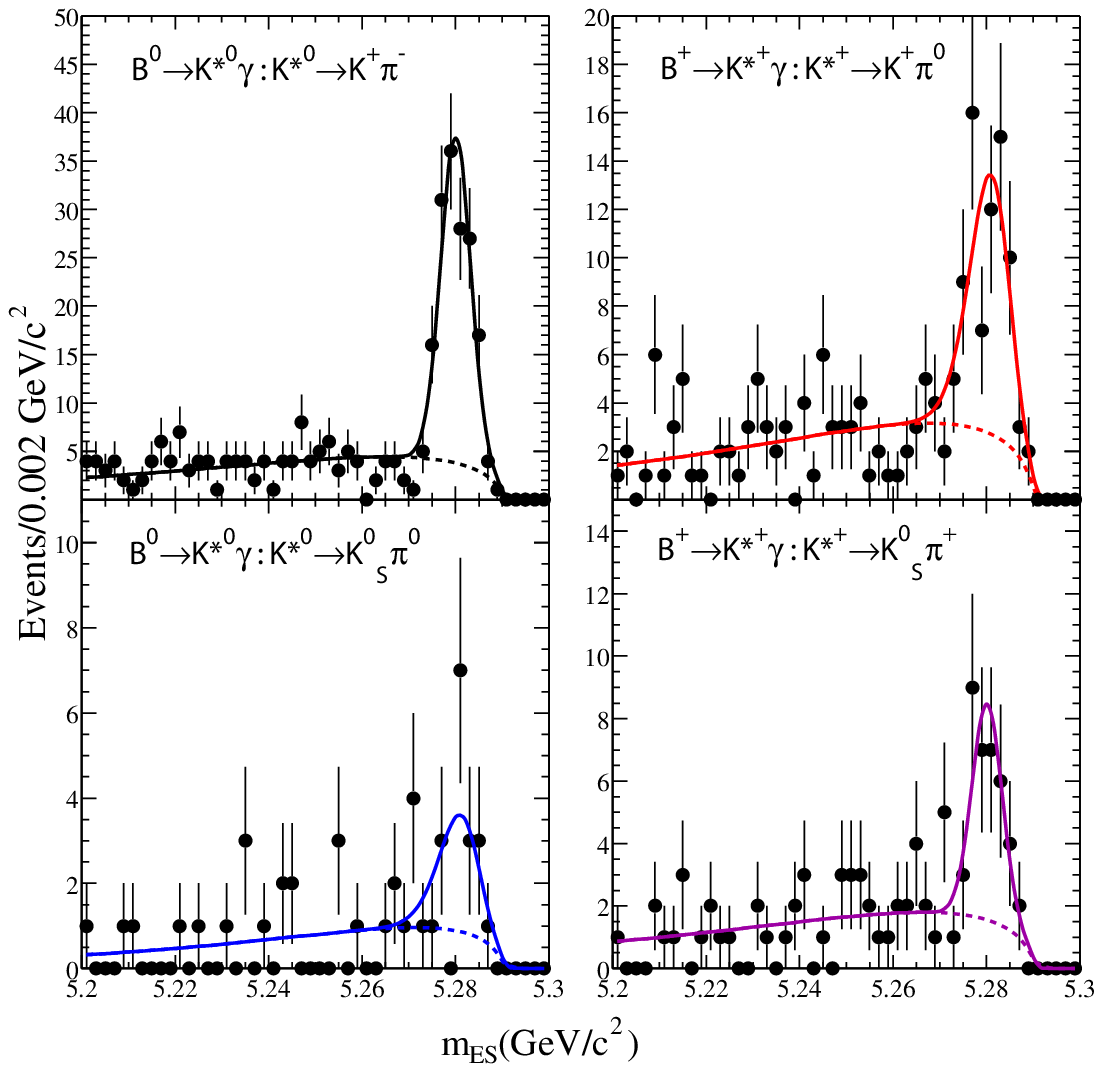}
\epsfysize=5cm\epsfbox{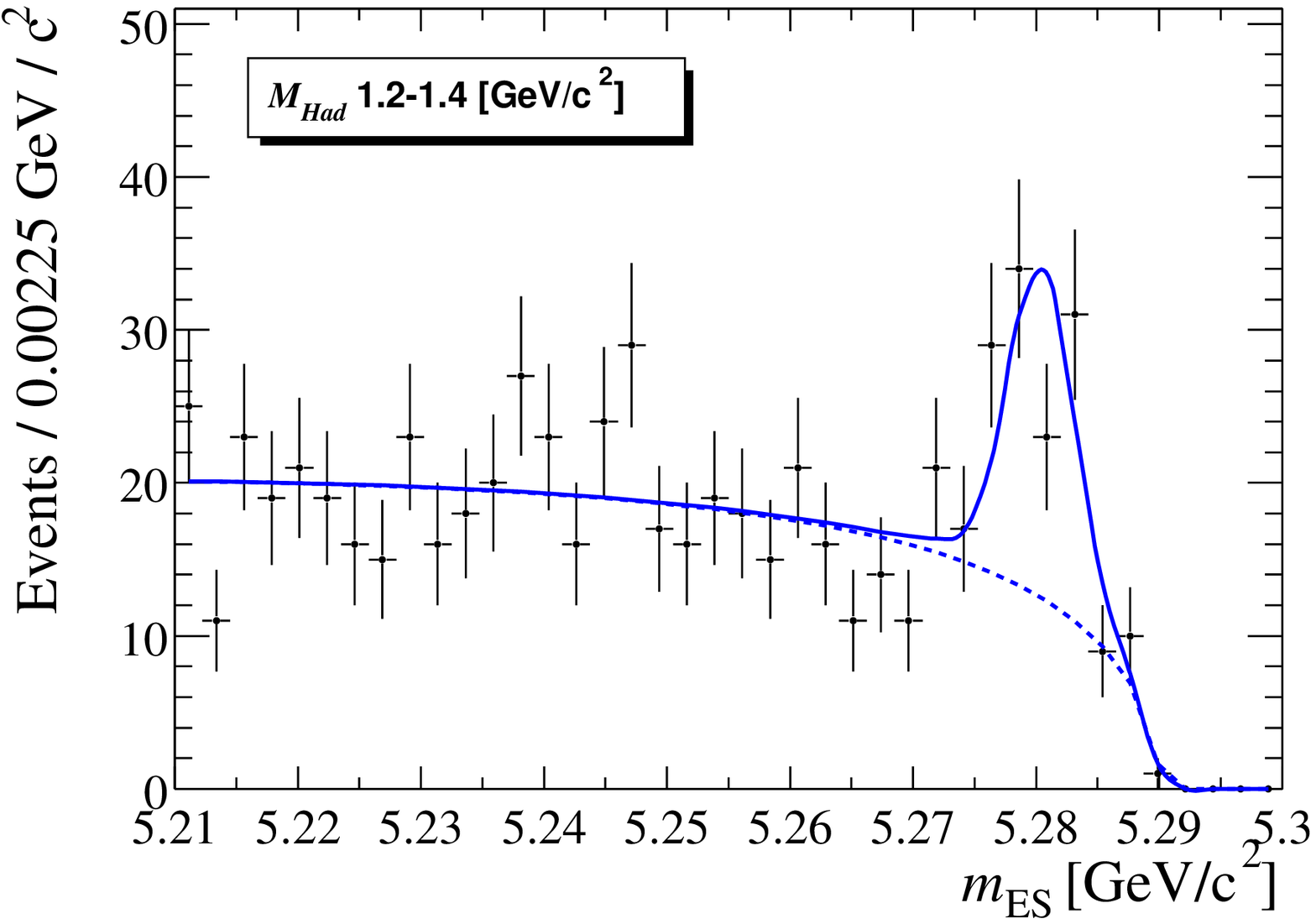}}
\caption{The $M_{ES}$ distribution for the decays $B \rightarrow K^{*} \gamma$
(left) and typical $M_{ES}$ distribution for the decays
$B \rightarrow s \gamma$ (right)}
\label{fig:ksgmes}
\end{figure}
The standard model prediction\cite{ksg} for this decay is about
$7.5 \pm 3.0 \times 10^{-5}$.
Event selection \cite{ksg} is based on finding a localized high energy photon
which does not come from $\pi^{0}$ and $\eta$ decay and does not have a
lateral energy profile in the calorimeter characteristic of neutral hadrons.
$\cal B$( $B^{0} \rightarrow K^{0*} \gamma$ ) = $4.23 \pm 0.40 \pm 0.22
\times 10^{-5}$,
%\newline
$\cal B$( $B^{+} \rightarrow K^{+*} \gamma$ ) = $3.83 \pm 0.62 \pm 0.22
\times 10^{-5}$
\section{$b \rightarrow s \gamma$}
\begin{figure}[htbp]
%\centerline{\epsfysize=4.25cm\epsfxsize=6cm\epsfbox{bsqBF.eps}}
%\centerline{\epsfysize=4.5cm\epsfxsize=6cm\epsfbox{bsqBF.eps}}
\centerline{\epsfysize=4.55cm\epsfbox{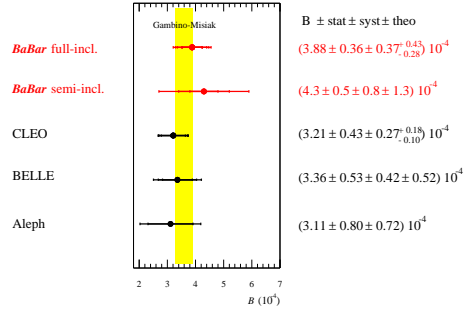}}
\caption{Comparison of results for the decay $b \rightarrow s \gamma$.}
\label{fig:bsgres}
\end{figure}
%\begin{figure}[htbp]
%\centerline{\epsfysize=3cm\epsfbox{bsqBF.eps}\epsfysize=3cm\epsfbox{utbrg.eps}}
%\caption{Comparison of results for the decay $b \rightarrow s \gamma$ (left)
%and Bounds in the $\rho$ $\eta$ plane from our limit on $B \rightarrow \rho \gamma$ (right)}.
%\label{fig:bsgres}
%\end{figure}
The expected branching ratio for $b \rightarrow s \gamma$ is in the
range\cite{kn}
$3.29 \pm 0.33 \times 10^{-4}$ to $3.60 \pm 0.33 \times 10^{-4}$.
Here we report on two different approaches:
a semi-inclusive method\cite{exclbsg}, in which one sums over $B$ decays,
or an inclusive analysis\cite{inbsg} where only the photon is measured.
The photon definition is similar to that used in the
$B \rightarrow K^{*} \gamma$ search.

The semi-inclusive method used twelve modes containing a $K^{+}$ or a
$K^{0}_{s}$ and up to 3 $\pi$ but only up to a single $\pi^{0}$.
Corrections are made for backgrounds and for undetected modes.
%Figure \ref{fig:ksgmes} (right)
In the inclusive method tight cuts were applied to ensure the presence of the
other B in the event and the photon energy restricted to the range
$2.1<E^{*}_{\gamma}<2.7$ GeV.  A 4$\pm$2\% correction is applied for contributions
coming from $b \rightarrow d \gamma$.  Results are summarized
in figure \ref{fig:bsgres}.
%\vspace{-0.5cm}
\section{$B \rightarrow \rho \gamma$}
The ratio of branching fractions
$B \rightarrow \rho \gamma$ / $B \rightarrow K^{*} \gamma$
constrains $V_{td}/V_{ts}$.  The present search\cite{radrho} included
$B^{0} \rightarrow \rho^{0} \gamma$, $B^{+} \rightarrow \rho^{+} \gamma$ and
$B^{0} \rightarrow \omega \gamma$.  No signal was found in a search of 84
million B pairs.  90\% c.l.~bounds on the decays
$\cal B$( $B^{0} \rightarrow \rho^{0} \gamma$ ) $< 1.4 \times 10^{-6}$,
$\cal B$( $B^{+} \rightarrow \rho^{+} \gamma$ ) $< 2.3 \times 10^{-6}$,
$\cal B$( $B^{0} \rightarrow \omega \gamma$ ) $< 1.2 \times 10^{-6}$ are
still a factor of 2-3 above standard model predicitions.

The combined limit assuming:\\
$\cal B$( $B \rightarrow \rho \gamma$ ) $\equiv$
2 $\cal B$( $B^{0} \rightarrow \rho^{0} \gamma$ ) =
2$\cal B$( $B^{0} \rightarrow \omega \gamma$ ) =
$\cal B$( $B^{+} \rightarrow \rho^{+} \gamma$ ) is\\
$\cal B$( $B \rightarrow \rho \gamma$ ) $< 1.9 \times 10^{-6}$.
Which sets a limit of $\left| \frac{V_{td}}{V_{ts}} \right| < 0.36$.
%$\left| \frac{V_{td}}{V_{ts}} \right| =
%\lambda \left| 1 - \rho -i \eta \right|$;\-\-
%$\left| \frac{V_{td}}{V_{ts}} \right| < 0.36$


\vspace{-0.75cm}
\begin{thebibliography}{99}
%CONF-02/023 hep-ex/0207082  SLAC-PUB-9323
\bibitem{kll} B.~Aubert {\it et al.}  [BABAR Collaboration],
``Evidence for the flavor changing neutral current
decays $B \to K l^{+} l^{-}$  and $B \to K^{*} l^{+} l^{-}$''
[arXiv:hep-ex/0207082].
%%CITATION = HEP-EX 0207082;%%
\bibitem{ksg} B.~Aubert {\it et al.}  [BABAR Collaboration],
``Measurement of $B \to K^{*} \gamma$ branching fractions and charge
asymmetries,'', Phys.\ Rev.\ Lett.\  {\bf 88}, 101805 (2002)
[arXiv:hep-ex/0110065].
\bibitem{kn} Kagan and Neubert, Euro.~Phys.~Jour.~C {\bf 7}, 5 (1999).
%CONF-02/026 hep-ex/0207076 BAD 464 and BAD 323
\bibitem{exclbsg}
B.~Aubert {\it et al.}  [BABAR Collaboration],
``$B \rightarrow s \gamma$ using a Sum of Exclusive Modes''
[arXiv:hep-ex/0207074].
%\bibitem{inbsg} hep-ph/9805303.
\bibitem{inbsg}
B.~Aubert {\it et al.}  [BABAR Collaboration],
``Determination of the branching fraction for inclusive decays
$B \rightarrow$ X$_{s}$ $\gamma$''
[arXiv:hep-ex/0207076].
\bibitem{gm} Gambino and Misiak, Nuc.~Phys. B{\bf 631}, 338 (2001).
%CONF-02/025 hep-ex/0207074 SLAC-PUB-9308
\bibitem{radrho}
B.~Aubert {\it et al.}  [BABAR Collaboration],
``Search for the exclusive radiative decays $B \rightarrow \rho \gamma$ and
$B^{0} \rightarrow \omega \gamma$'',
[arXiv:hep-ex/0207073].
%CONF-02/024 hep-ex/0207073  SLAC-PUB-9319
\end{thebibliography}
\end{document}